\documentclass[sigconf,nonacm]{acmart}
\usepackage{booktabs}
\usepackage{xcolor}

\newcommand{\NTasks}{800}

\newcommand{\CorrTauK}{-0.03}

\newcommand{\SpecMatchHaikuClosed}{0.946}
\newcommand{\CompileOKHaikuOpen}{0.948}

\newcommand{\SilentHaikuOpen}{0.806}

\newcommand{\ActionMatchGptClosed}{0.582}

\newcommand{\SpecMatchGptClosed}{0.938}
\newcommand{\CompileOKGptOpen}{0.955}
\newcommand{\ActionMatchGptOpen}{0.544}

\newcommand{\SilentGptOpen}{0.729}
\newcommand{\ExactGptOpen}{0.087}
\newcommand{\ESmedGptOpen}{442}

\newcommand{\SpecMatchQwenLClosed}{0.935}
\newcommand{\CompileOKQwenLOpen}{0.601}
\newcommand{\ActionMatchQwenLOpen}{0.172}

\newcommand{\CompileOKQwenSClosed}{0.488}

\newcommand{\SpecOneBehaviourWrong}{0.792}
\newcommand{\NSpecOne}{1767}
\newcommand{\SpecOneWrongGpt}{0.666}

\newcommand{\AMQoneGptOpen}{0.674}
\newcommand{\AMQfiveGptOpen}{0.473}

\newcommand{\SpecOneWrongHaiku}{0.791}

\newcommand{\AMQoneHaikuOpen}{0.515}
\newcommand{\AMQfiveHaikuOpen}{0.315}

\newcommand{\SpecOneWrongQwenL}{0.934}

\newcommand{\BestAMOpen}{0.544}

\newcommand{\BestExactOpen}{0.087}

\newcommand{\CompileLenientQwenLOpen}{0.706}
\newcommand{\AMLenientQwenLOpen}{0.169}

\newcommand{\CompileLenientQwenSClosed}{0.713}

\newcommand{\SilentEightyMin}{0.370}
\newcommand{\SilentEightyMax}{0.741}

\newcommand{\SilentNinetyNineMin}{0.372}
\newcommand{\SilentNinetyNineMax}{0.884}
\newcommand{\SilentGptOpenEighty}{0.655}
\newcommand{\SilentGptOpenNinetyNine}{0.840}
\newcommand{\RNetMed}{-35.1}
\newcommand{\RGrossMed}{-0.6}
\newcommand{\ShareNetPos}{8.9}
\newcommand{\ShareGrossPos}{47.8}
\newcommand{\RegOpenTauGivenR}{-0.063}
\newcommand{\RegClosedTauGivenR}{-0.086}
\newcommand{\RegOpenRonly}{-0.11}
\newcommand{\RegOpenRonlyP}{3e-05}

\newcommand{\AMSubOpus}{0.889}
\newcommand{\ExactSubOpus}{0.575}
\newcommand{\SilentSubOpus}{0.275}
\newcommand{\AMQoneSubOpus}{0.864}
\newcommand{\AMQfiveSubOpus}{0.884}

\newcommand{\AMSubGpt}{0.575}
\newcommand{\ExactSubGpt}{0.085}

\newcommand{\NSub}{200}

\newcommand{\JudgePassBadOpus}{55}
\newcommand{\JudgePassOkOpus}{145}

\newcommand{\JudgeSilentShareOpus}{0.275}

\newcommand{\JudgePassBadGpt}{120}
\newcommand{\JudgePassOkGpt}{43}

\newcommand{\JudgeSilentShareGpt}{0.736}

\newcommand{\JudgeMissing}{26}
\newcommand{\NegCtlN}{7}
\newcommand{\NegCtlRejected}{7}
\newcommand{\RegOpenTauOnlyTau}{-0.065}
\newcommand{\RegOpenTauOnlyTauSE}{0.013}

\newcommand{\RegOpenTauKTau}{-0.066}

\newcommand{\RegOpenTauKFeTau}{-0.041}

\newcommand{\RegOpenTauKFeTauP}{0.014}

\newcommand{\RegOpenTauKFeControlsTau}{-0.033}

\newcommand{\RegOpenTauKFeControlsTauP}{0.049}

\newcommand{\RegClosedTauOnlyTau}{-0.088}
\newcommand{\RegClosedTauOnlyTauSE}{0.013}

\newcommand{\RegClosedTauKTau}{-0.088}

\newcommand{\RegClosedTauKFeTau}{-0.058}

\newcommand{\RegClosedTauKFeTauP}{0.00023}

\newcommand{\RegClosedTauKFeControlsTau}{-0.063}

\newcommand{\RegClosedTauKFeControlsTauP}{3.7e-05}

\newcommand{\NBlocksSignal}{6}
\newcommand{\NBlocksFilter}{2}
\newcommand{\NBlocksSizing}{3}
\newcommand{\NBlocksRisk}{9}
\newcommand{\NPool}{15000}
\newcommand{\NPoolKept}{14393}
\newcommand{\NFewTrades}{607}
\newcommand{\NAlwaysFlat}{63}
\newcommand{\RMedBinA}{-0.37}
\newcommand{\RMedBinB}{-0.37}
\newcommand{\RMedBinC}{-0.37}
\newcommand{\RMedBinD}{-0.34}
\newcommand{\RMedBinE}{-0.34}
\newcommand{\NValidPrompts}{796}
\newcommand{\NRoundTripOK}{780}
\newcommand{\NRoundTripNA}{20}
\newcommand{\MedWords}{96}
\newcommand{\MedJargon}{11}

\begin{document}
\title{MintEval: Do LLMs Implement the Trading Strategy You Asked For?}
\subtitle{A Behavioural-Equivalence Benchmark for Natural-Language-to-Strategy Code}

\author{Varstern (Yifan) Wang}
\affiliation{%
  \institution{Fudan University}
  \city{Shanghai}
  \country{China}}
\email{wag25528@gmail.com}

\author{Siyu Wang}
\affiliation{%
  \institution{Fudan University}
  \city{Shanghai}
  \country{China}}
\email{24300680134@m.fudan.edu.cn}

\author{Yuecheng He}
\affiliation{%
  \institution{Fudan University}
  \city{Shanghai}
  \country{China}}
\email{yche24@m.fudan.edu.cn}

\renewcommand{\shortauthors}{Wang et al.}

\begin{abstract}
Large language models are moving from producing trading \emph{signals} to writing the code that
executes them. The failure mode of the second role is silent: generated code runs, a backtest
plots, yet the risk logic that the trader described is not the logic being executed. Existing
code benchmarks test functional correctness on unit tests and finance benchmarks test forecasting;
neither measures whether an implementation \emph{behaves} like the strategy that was asked for.
We introduce MintEval, a benchmark in which reference strategies are generated programmatically
from a library of composable building blocks, back-translated into colloquial trader instructions,
and re-implemented by the model under test. Generated and reference programs are executed bar by
bar on identical market data and frictions, and compared on their actions rather than on code
similarity or profit: alpha is differenced away. MintEval v0 contains \NTasks{} tasks on BTCUSDT
15-minute data, stratified by an execution-measured state-span complexity $\tau$ that is decoupled
from description length. Low-cost models reach a mean ActionMatch of at most \BestAMOpen{} and reproduce at most
\BestExactOpen{} of tasks exactly; on a stratified subset of \NSub{} tasks a frontier model (Claude
Opus 5.5) reaches \AMSubOpus{} and reproduces \ExactSubOpus{} exactly, yet still fails silently on
\SilentSubOpus{} of tasks. Given a menu of building blocks, models identify the strategy almost
perfectly, yet \SpecOneBehaviourWrong{} of the implementations whose specification was read correctly
diverge on more than 10\% of active bars. The LLM judge of a recent strategy-generation benchmark,
applied verbatim, accepts every one of these silent failures.
\end{abstract}

\ccsdesc[500]{Software and its engineering~Software verification and validation}
\ccsdesc[300]{Computing methodologies~Natural language processing}
\ccsdesc[100]{Applied computing~Forecasting}
\keywords{LLM evaluation, code generation, benchmark, algorithmic trading, behavioural equivalence, stateful programs}

\maketitle

\section{Introduction}
LLMs are increasingly asked not only for a market view \cite{lopezlira2023chatgpt,wu2023bloomberggpt,yang2023fingpt}
but for the code that turns a trader's intent into orders. In that role an error is not a bad
forecast but an implementation shortfall \cite{perold1988implementation} between the strategy that
was intended and the one that runs. Such errors are \emph{silent}: the code compiles, the backtest
completes and produces a plausible equity curve, and nothing signals that, say, the breakeven stop
never moves or the cooldown resets every bar. Operational failures of trading code are costly
\cite{sec2013knight}, and backtests are already a fragile basis for decisions \cite{bailey2014pseudo,bailey2016pbo}.

Code benchmarks such as HumanEval \cite{chen2021humaneval}, MBPP \cite{austin2021mbpp},
EvalPlus \cite{liu2023evalplus}, DS-1000 \cite{lai2023ds1000}, BigCodeBench \cite{zhuo2024bigcodebench}
and SWE-bench \cite{jimenez2024swebench} score functional correctness against unit tests that a
human wrote for a precise specification. Trading strategies differ in three ways that make that
template insufficient: (i) they are \emph{stateful} over thousands of steps, and the hard part is
the persistent state (an entry price, a ratcheted stop, a cooldown clock); (ii) they have
\emph{zero tolerance}: a single mis-timed exit changes every later decision; (iii) the
specification arrives as \emph{colloquial, jargon-laden} language, not as a docstring.

\textbf{Contributions.} (1) A task definition, $\Phi: q_s \mapsto \hat c_s$, scored by bar-level
\emph{behavioural equivalence} with a reference program under identical data and frictions.
(2) A metric suite led by ActionMatch on active bars. (3) A generator of contamination-free tasks
with two separated complexity axes: description length $K$ (generation side) and state span
$\tau$ and register count $R$ (measured by instrumenting execution). (4) Results for four models in an open setting (instruction + interface only) and a closed
diagnostic setting (with a menu of building blocks), separating understanding from implementation.

\section{Task}
A task $s$ consists of a reference program $c_s$, a colloquial instruction $q_s$ and fixed
market data $D$ with frictions $F$. The model under test receives $q_s$ and an interface
document (engine semantics, the indicator library, and desk conventions such as ``entry
price is the close of the signal bar'') and returns $\hat c_s$. Both programs implement
\texttt{strategy(hist, state, pos, ind)}, are called at every bar close, and return a target
position in $[-1,1]$ plus optional stop and take-profit levels for the next bar. We call
$\hat c_s$ \emph{behaviourally equivalent} to $c_s$ on $(D,F)$ if their quantised target
sequences coincide on every bar. Equivalence is a property of behaviour, not of code: two
programs with different structure are equivalent if they act identically, and two programs
that differ by one comparison operator are not.

The engine enforces causality physically: bar $t$ is copied into the visible buffers right
before the call, indicators are revealed bar by bar, and any access to $t{+}1$ raises an error
that cannot be swallowed. State is a flat dictionary of scalars whose reads and writes are logged.
Generated code runs in a subprocess sandbox (imports limited to \texttt{math}/\texttt{numpy}, no
I/O, 10\,s budget).

\section{Metric Suite}
\textbf{ActionMatch} (primary) is the fraction of \emph{active} bars on which the two quantised
targets are equal, where a bar is active if either program holds or targets a non-zero position.
Restricting to active bars matters: flat periods are trivially matched and would otherwise
inflate every score. \textbf{TradeF1} matches round trips exactly (direction, entry bar, exit bar).
$|\mathrm{ES}|$ is the absolute difference in total return in basis points of initial equity; we
take the absolute value because an implementation error that happens to make money is still an
error, and signed errors would cancel across tasks. We also report $\Delta$Sharpe and $\Delta$MaxDD.
A run that raises, times out or looks ahead is \textbf{CompileOK}$=0$; its other metrics are left
missing rather than imputed. In the closed setting, \textbf{SpecMatch} scores the model's
structured reading of $q_s$ slot by slot. A \emph{silent failure} is a run with CompileOK$=1$
and ActionMatch$<0.9$. The threshold is a convention, not a cliff: at 0.8 the silent share
ranges from \SilentEightyMin{} to \SilentEightyMax{} across models and settings, and at 0.99 from
\SilentNinetyNineMin{} to \SilentNinetyNineMax{} (GPT-5.4-mini open: \SilentGptOpenEighty{} at 0.8,
\SilentGptOpen{} at 0.9, \SilentGptOpenNinetyNine{} at 0.99).

\section{Benchmark Construction}
\textbf{Building blocks.} The library has \NBlocksSignal{} signal, \NBlocksFilter{} filter,
\NBlocksSizing{} sizing and \NBlocksRisk{} risk blocks, each
with a discrete parameter grid and a precise English template. Programs combine one signal,
a direction, a filter, a sizing rule and 1--3 risk layers; hard combinations include moving
boxes, scaling out, pyramiding, multi-timeframe filters, breakout-then-retest state machines and
risk layers that read each other's state (a trailing stop that arms only after breakeven).

\textbf{Two complexity axes.} $K$ (bits) is the description length of the program in the
generator's grammar. $\tau_{\max}$ and $\tau_{p90}$ are the maximum and 90th percentile, over all
value reads of the state dictionary during the reference backtest, of the number of bars since
the key was last written; $R$ counts keys with at least one cross-bar read. Across the task set
$\mathrm{corr}(\log\tau_{\max},K)=\CorrTauK$. Several blocks vary $\tau$ through a parameter
alone (time stop, cooldown, retest window), which lets us estimate the effect of $\tau$ within a
block via block fixed effects.

\textbf{Selection.} We sample \NPool{} candidate programs and keep the \NPoolKept{} that make at least
20 round trips, are not flat throughout and do not lose half their equity in the first 5\% of bars
(\NFewTrades{} removed for too few trades, \NAlwaysFlat{} for never trading). Because programs with
short state spans trade more often and therefore pay more fees, $\tau$ and reference return are
confounded in the raw pool. We therefore stratify jointly: within each $\tau_{\max}$ quintile we draw
32 tasks from each of five equal-width bands of reference return, giving \NTasks{} tasks with median
reference return \RMedBinA, \RMedBinB, \RMedBinC, \RMedBinD{} and \RMedBinE{} across the five quintiles.

\textbf{Why losing references do not undermine the measurement.} The median reference program
loses \RNetMed\% over the two years and only \ShareNetPos\% are profitable after costs. This is
friction, not a negative edge: re-running the same programs with zero fees and slippage gives a
median of \RGrossMed\% with \ShareGrossPos\% profitable, i.e.\ randomly assembled rules with no edge
that trade frequently at 6\,bp per fill. The measurement does not depend on the sign of P\&L:
ActionMatch and TradeF1 compare actions, and $|$ES$|$ compares two programs under identical
frictions, so a faithful re-implementation of a losing strategy scores perfectly and an unfaithful
one does not. Empirically, ActionMatch is only weakly related to reference return, and in the
direction opposite to the concern: references that lose more are slightly \emph{easier} to match
(\RegOpenRonly{} ActionMatch per 100\% of return, open setting, $p=\RegOpenRonlyP$), and the $\tau$
coefficient is unchanged when $R_{\text{bench}}$ is added as a control (\RegOpenTauGivenR{} open,
\RegClosedTauGivenR{} closed, vs.\ \RegOpenTauOnlyTau{} and \RegClosedTauOnlyTau{} without it). A
benchmark whose references are realistic trading candidates is nevertheless more convincing for a
finance audience; v1 lowers frictions to 2--3\,bp and adds 1-hour bars so that the reference return
distribution straddles zero.

\textbf{Back-translation.} An LLM translator (a third vendor, not among the models under test)
rewrites each program, given its precise template description and source, as a 40--120 word desk
message in one of eight rotating voices. Deterministic gates require every parameter value to appear,
at least three expressions from a slang dictionary (drafted with LLM help, under human review), no
code-like tokens, and the conventional phrasing for formula-level rules; failures are fed back for up
to six attempts. A round-trip reader from a fourth vendor then recovers the block-level spec from the
message; a mismatch is fed back as well. \NValidPrompts{} of \NTasks{} messages pass all gates
(\NRoundTripOK{} round-trip verified, \NRoundTripNA{} unverifiable because the reader failed); median
length \MedWords{} words with \MedJargon{} slang hits. The reader sees the block menu, so it verifies
identification rather than sufficiency; two ambiguities it missed during calibration were resolved
with explicit desk conventions (\S\ref{sec:limits}).

\textbf{Contamination.} Tasks are sampled from a combinatorial space at build time; neither the
instruction nor the reference program exists on the web.

\section{Preliminary Results}
\begin{table*}[t]\centering\small
\caption{Leaderboard on \NTasks{} tasks. ActionMatch, TradeF1 and $|$ES$|$ are over compiled runs;
Silent = share of all tasks that compile but have ActionMatch $<0.9$.}
\label{tab:leader}
\begin{tabular}{llrrrrrr}
\toprule
Model & Set. & CompileOK & SpecMatch & ActionMatch & TradeF1 & $|$ES$|$ med (bp) & Silent \\
\midrule
gpt-5.4-mini & open & 0.955 & -- & 0.544 & 0.469 & 442 & 0.729 \\
claude-haiku & open & 0.948 & -- & 0.400 & 0.350 & 752 & 0.806 \\
qwen2.5-coder-32b & open & 0.601 & -- & 0.172 & 0.191 & 1646 & 0.586 \\
qwen2.5-coder-7b & open & 0.374 & -- & 0.081 & 0.043 & 2252 & 0.372 \\
\midrule
gpt-5.4-mini & closed & 0.984 & 0.938 & 0.582 & 0.541 & 362 & 0.665 \\
claude-haiku & closed & 0.956 & 0.946 & 0.476 & 0.403 & 613 & 0.755 \\
qwen2.5-coder-32b & closed & 0.613 & 0.935 & 0.240 & 0.185 & 1360 & 0.576 \\
qwen2.5-coder-7b & closed & 0.488 & 0.741 & 0.058 & 0.021 & 2781 & 0.487 \\
\bottomrule
\end{tabular}

\end{table*}

\begin{figure}[t]\centering
\includegraphics[width=\linewidth]{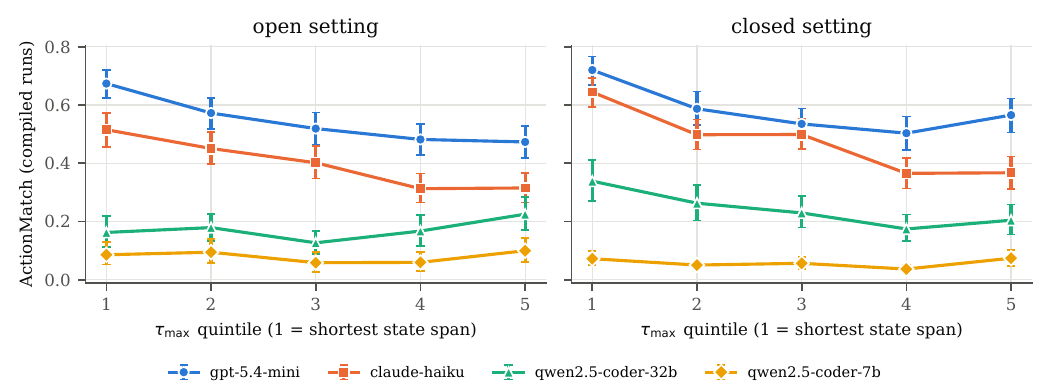}
\caption{ActionMatch by $\tau_{\max}$ quintile (mean, 95\% bootstrap CI).}
\Description{Two panels, open and closed setting, showing mean ActionMatch with confidence
intervals for each model across five quintiles of state span.}
\label{fig:tau}
\end{figure}

\begin{figure}[t]\centering
\includegraphics[width=\linewidth]{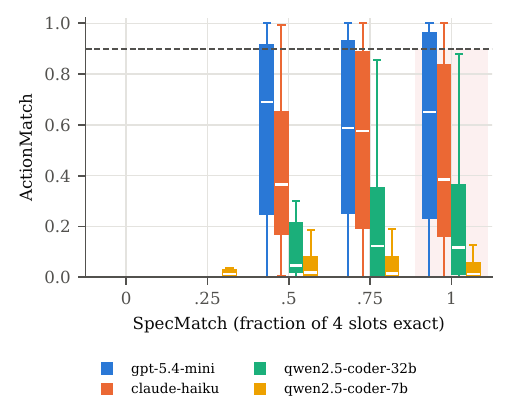}
\caption{Closed setting: SpecMatch vs ActionMatch per task. Shaded: specification fully correct
but behaviour diverges.}
\Description{Scatter plot of ActionMatch against SpecMatch for each task and model; points with
SpecMatch equal to one and ActionMatch below 0.9 are highlighted.}
\label{fig:spec}
\end{figure}

\textbf{Models and setup.} Two low-cost closed models from different vendors (GPT-5.4-mini,
Claude Haiku) and two open-weight coders (Qwen2.5-Coder 32B and 7B, served locally with vLLM in
batch-invariant mode); temperature 0, one sample per task and setting.

\textbf{Low-cost models are far from saturation} (Table~\ref{tab:leader}). In the open setting the
strongest of them reaches ActionMatch \ActionMatchGptOpen{} on compiled runs and reproduces \ExactGptOpen{} of
tasks exactly; the median $|$ES$|$ is \ESmedGptOpen{}\,bp of initial equity. Compilation is not the
bottleneck for the closed models (CompileOK \CompileOKGptOpen{} and \CompileOKHaikuOpen{}): most of
their failures are silent, \SilentGptOpen{} and \SilentHaikuOpen{} of all tasks.

\textbf{A frontier model narrows the gap but does not close it} (Table~\ref{tab:frontier}). Budget
allowed one frontier model on a subset of \NSub{} tasks (40 per $\tau$ quintile, open setting). Claude
Opus 5.5 compiles every task, reaches ActionMatch \AMSubOpus{} and reproduces \ExactSubOpus{} of tasks
exactly, against \AMSubGpt{} and \ExactSubGpt{} for GPT-5.4-mini on the same tasks; but
\SilentSubOpus{} of its tasks are still silent failures, and unlike the weaker models it shows no
$\tau$ gradient (\AMQoneSubOpus{} in the first quintile, \AMQfiveSubOpus{} in the last). At the
frontier, the residual errors are not concentrated in long-state tasks; v1 needs harder tasks and
frontier models from other families (see Limitations on authorship).

\begin{table}[t]\centering\scriptsize\setlength{\tabcolsep}{3pt}
\caption{Same \NSub{} tasks (40 per $\tau_{\max}$ quintile), open setting. ActionMatch and the
quintile columns are over compiled runs; Exact and Silent are shares of all tasks.}
\label{tab:frontier}
\begin{tabular}{lrrrrrrrrr}
\toprule
Model & CompileOK & ActionMatch & Exact & Silent & $\tau$1 & $\tau$2 & $\tau$3 & $\tau$4 & $\tau$5 \\
\midrule
claude-opus-5.5 & 1.000 & 0.889 & 0.575 & 0.275 & 0.86 & 0.91 & 0.92 & 0.87 & 0.88 \\
gpt-5.4-mini & 0.945 & 0.575 & 0.085 & 0.710 & 0.73 & 0.64 & 0.57 & 0.52 & 0.42 \\
claude-haiku & 0.955 & 0.420 & 0.040 & 0.820 & 0.53 & 0.55 & 0.49 & 0.31 & 0.22 \\
qwen2.5-coder-32b & 0.635 & 0.181 & 0.005 & 0.615 & 0.08 & 0.22 & 0.15 & 0.22 & 0.22 \\
qwen2.5-coder-7b & 0.330 & 0.076 & 0.000 & 0.330 & 0.07 & 0.09 & 0.07 & 0.04 & 0.09 \\
\bottomrule
\end{tabular}

\end{table}

\textbf{Understanding is not the bottleneck} (Fig.~\ref{fig:spec}). With the block menu, GPT-5.4-mini,
Claude Haiku and Qwen-32B recover the specification with SpecMatch \SpecMatchGptClosed,
\SpecMatchHaikuClosed{} and \SpecMatchQwenLClosed. Yet among the \NSpecOne{} compiled runs whose
specification is fully correct, \SpecOneBehaviourWrong{} have ActionMatch below 0.9
(GPT-5.4-mini \SpecOneWrongGpt, Haiku \SpecOneWrongHaiku, Qwen-32B \SpecOneWrongQwenL). The closed
setting improves ActionMatch only modestly over the open one (e.g.\ \ActionMatchGptOpen{} $\to$
\ActionMatchGptClosed), so most of the gap is in implementing persistent logic, not in reading
the instruction.

\textbf{State span matters} (Fig.~\ref{fig:tau}). For the two stronger models, ActionMatch falls
from the shortest to the longest $\tau_{\max}$ quintile (GPT-5.4-mini open \AMQoneGptOpen{} $\to$
\AMQfiveGptOpen; Haiku open \AMQoneHaikuOpen{} $\to$ \AMQfiveHaikuOpen). Pooling models, an OLS of
ActionMatch on $\log_{10}\tau_{\max}$ with model dummies and task-clustered errors gives
\RegOpenTauOnlyTau{} (s.e.\ \RegOpenTauOnlyTauSE) per decade in the open setting and
\RegClosedTauOnlyTau{} (\RegClosedTauOnlyTauSE) in the closed one. Because the generator makes
$\tau$ nearly orthogonal to description length ($\mathrm{corr}=\CorrTauK$), adding $K$ leaves the
estimate unchanged (\RegOpenTauKTau, \RegClosedTauKTau). With signal and risk-block fixed effects and
register count it shrinks to \RegOpenTauKFeTau{} ($p=\RegOpenTauKFeTauP$) and \RegClosedTauKFeTau{}
($p=\RegClosedTauKFeTauP$); adding McCabe complexity and log trade count leaves \RegOpenTauKFeControlsTau{}
($p=\RegOpenTauKFeControlsTauP$, borderline) and \RegClosedTauKFeControlsTau{}
($p=\RegClosedTauKFeControlsTauP$). The within-block effect is therefore smaller than the raw
gradient but does not vanish; the open-setting estimate under full controls is borderline and
needs the larger v1 set. The Qwen models are near the floor, where no gradient can show.

\textbf{What goes wrong.} For the open-weight models, failures that raise are dominated by
interface misuse (Qwen-7B indexes the history object directly), state-type violations (storing
\texttt{None}, which the interface forbids) and look-ahead: indexing a higher-timeframe series
with the 15-minute bar index (e.g.\ \texttt{htf.close[t // 4]}), which reads the candle still in
progress. In an engine without physical causality enforcement this silently leaks future data
into the backtest. Silent failures we inspected include breakout levels computed over a window that
contains the current bar (so the breakout can never trigger), stops re-anchored to the current
close on every bar instead of to the entry, and HTF filters evaluated on the wrong candle. A
systematic taxonomy is left to v1.

\subsection*{A code-reading judge does not see silent failures}\label{sec:judge}
We apply the QuantCode-Bench judge \cite{khoroshilov2026quantcode} verbatim (prompt and rating parser
from its public repository; temperature 0, thinking off; the judge model is Claude Sonnet 4, the
default in that repository) to every compiled open-setting run of Opus 5.5 and GPT-5.4-mini on the
\NSub{}-task subset, giving it our instruction $q_s$ and the generated code.

\begin{table}[h]\centering\small
\caption{QuantCode-Bench judge vs.\ behavioural equivalence (ActionMatch $\geq 0.9$).}
\label{tab:judge}
\begin{tabular}{llrr}
\toprule
Model & & Judge pass & Judge fail\\
\midrule
Opus 5.5 & ActionMatch $\geq 0.9$ & \JudgePassOkOpus & 0\\
 & ActionMatch $< 0.9$ & \textbf{\JudgePassBadOpus} & 0\\
GPT-5.4-mini & ActionMatch $\geq 0.9$ & \JudgePassOkGpt & 0\\
 & ActionMatch $< 0.9$ & \textbf{\JudgePassBadGpt} & 0\\
\bottomrule
\end{tabular}
\end{table}

The judge passes every judged run (Table~\ref{tab:judge}): all \JudgePassBadOpus{} silent failures
of Opus 5.5 (\JudgeSilentShareOpus{} of its passes) and all \JudgePassBadGpt{} of GPT-5.4-mini
(\JudgeSilentShareGpt{} of its passes; \JudgeMissing{} of its runs could not be judged because our
API budget ran out). The judge is not simply permissive: in a negative control that pairs an
instruction with code written for a different task using a different signal family, it rejected
\NegCtlRejected{} of \NegCtlN. It detects the wrong strategy; it cannot detect the right strategy
implemented wrongly. This is the gap behavioural equivalence closes, and it applies to any
evaluation whose final check reads code rather than executing it against a reference.

\textbf{Sensitivity to the state rule.} Re-running every program that failed only because it stored
\texttt{None} or a string in the state dictionary under a lenient engine raises CompileOK for
Qwen-32B from \CompileOKQwenLOpen{} to \CompileLenientQwenLOpen{} (open) and for Qwen-7B from
\CompileOKQwenSClosed{} to \CompileLenientQwenSClosed{} (closed), but leaves ActionMatch on compiled
runs essentially unchanged (Qwen-32B open \ActionMatchQwenLOpen{} vs.\ \AMLenientQwenLOpen): the
rescued programs are mostly silent failures. The strict rule moves failures between the loud and
the silent column; it does not drive the conclusions.

\section{Related Work}
\textbf{Code generation benchmarks} score functional correctness on unit tests
\cite{chen2021humaneval,austin2021mbpp,liu2023evalplus,lai2023ds1000,zhuo2024bigcodebench} or on
repository tests \cite{jimenez2024swebench}; LiveCodeBench addresses contamination by collecting
problems after model cut-offs \cite{jain2024livecodebench}. MintEval instead generates tasks
combinatorially and judges behaviour over thousands of stateful steps against a reference, which
makes silent divergence measurable. \textbf{Closest to our work}, QuantCode-Bench \cite{khoroshilov2026quantcode} collects 400 real strategy
descriptions (Reddit, TradingView, StackExchange, GitHub) and checks generated Backtrader code for
compilation, backtest execution, the presence of trades, and semantic alignment judged by an LLM;
the best models reach 70--76\% judge pass in a single turn and 95--98\% with iterative repair. Its
tasks are real and its model coverage broad, both stronger than ours. The two benchmarks answer
complementary questions: QuantCode-Bench asks whether a model can produce an executable strategy
that plausibly matches a real description; MintEval asks whether the produced strategy
\emph{behaves} as described, which requires a reference program and lets us vary the mechanism.
In their pooled single-turn breakdown only 2.7\% of attempts are rejected by the judge; a
code-reading judge cannot observe a trailing stop that is re-anchored every bar. We measure this
gap directly in Section~\ref{sec:judge}. \textbf{Code hallucination} studies catalogue incorrect but
plausible code \cite{tian2024codehalu,zhang2024hallucode}; we quantify one consequential form of it.
\textbf{LLMs in finance} have been studied mainly as forecasters and domain language models
\cite{lopezlira2023chatgpt,wu2023bloomberggpt,yang2023fingpt}. Our metric borrows the logic of
implementation shortfall \cite{perold1988implementation}: the gap between the intended and the
executed strategy, with alpha differenced away. Backtest overfitting \cite{bailey2014pseudo,bailey2016pbo}
and operational failures \cite{sec2013knight} motivate treating implementation fidelity as a
first-class risk.

\section{Limitations}\label{sec:limits}
MintEval is a \emph{diagnostic} benchmark: controlled generation lets us vary $\tau$, $K$ and block
type independently, which a benchmark of naturally occurring tasks cannot do, but it does not
sample the distribution of real strategies. A real subset (strategies ported from open-source
frameworks and exchange product specifications, with human-written intents and human-verified
reference ports) is under construction; its role is to test whether model rankings and the $\tau$
effect transfer. v0 is small: one instrument (BTCUSDT), one data window, \NTasks{} tasks and four
models with one sample each. A human baseline and a human-rewritten instruction subset are in progress; until
then, instructions are LLM-written and may be friendlier to LLMs than real traders' messages, and
the slang dictionary is still under human review. Latency is deliberately excluded: fidelity and
latency are orthogonal goals. The unit of observation is the bar-level action and the complete
round trip; we do not model intra-position dynamics such as fills inside a bar beyond stop/take
priority. The reference library, interface document and desk conventions were written with the help of
Claude Opus 5.5, the same family as our frontier model; shared conventions may favour it, and
replicating the frontier result with models from other families is a priority for v1. Finally, a mismatch can stem from an ambiguous instruction rather than from the model;
the round-trip reader bounds but does not eliminate this.

\section*{Ethics Statement}
MintEval evaluates implementation fidelity, not profitability; no strategy in it is
investment advice, and most reference strategies lose money after costs. Market data are public
(data.binance.vision). No personal data are used. Model-generated code runs in a sandbox
without network or file access.

The benchmark suite, prompt templates, and evaluation harness are available at \url{https://github.com/spearmintai/minteval}.

\appendix
\section{$\tau_{\max}$ distribution}
\begin{figure}[h]\centering
\includegraphics[width=0.8\linewidth]{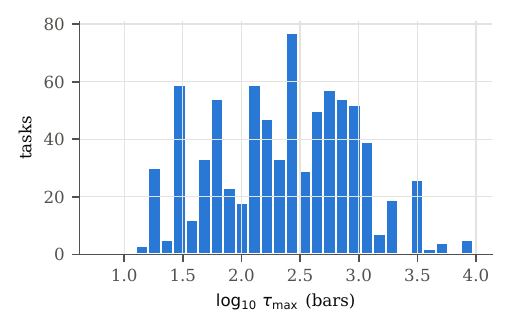}
\caption{Histogram of $\log_{10}\tau_{\max}$ over the task set.}
\Description{Histogram of the base-10 logarithm of maximum state span across tasks.}
\end{figure}
\end{document}